\documentclass[epjST]{svjour}
\usepackage{amssymb}
\usepackage{amsmath}
\usepackage{graphics}
\usepackage[colorlinks=true]{hyperref}

\hypersetup{urlcolor=blue, citecolor=red}

\begin{document}

\title{Permutation complexity of interacting dynamical systems}
\author{Roberto Monetti\inst{1}\fnmsep\thanks{%
monetti@mpe.mpg.de} \and Jos\'e Mar\'{\i}a Amig\'o\inst{2} \and Thomas
Aschenbrenner\inst{1} \and Wolfram Bunk\inst{1}}
\institute{Max-Planck-Institut f\"ur extraterrestrische Physik, Giessenbachstr. 1,
85748, Garching, Germany \and Centro de Investigaci\'on Operativa,
Universidad Miguel Hernandez, Avda. de la Universidad s/n, 03202 Elche, Spain}

\abstract{
The coupling complexity index is an information measure introduced within the framework of ordinal symbolic
dynamics. This index is used to characterize the complexity of the
relationship between dynamical system components. In this work, we clarify the meaning of the coupling complexity
by discussing in detail some cases leading to extreme values, and present examples using synthetic data to describe
its properties. We also generalize the coupling complexity index to the multivariate case and derive a number
of important properties by exploiting the structure of the symmetric group.
The applicability of this index to the multivariate case is demonstrated with a real-world data example. Finally,
we define the coupling complexity rate of random and deterministic time series.
Some formal results about the multivariate coupling complexity index have been collected in an Appendix.
}

\maketitle

\section{Introduction}

The characterization of complex dynamical systems is a relevant topic
arising in different fields of research. This kind of systems are often
composed of a large number of interacting components, thus the dynamical
behavior may depend on many degrees of freedom. Complex systems display a
variety of interesting phenomena, e.g. synchronization \cite{Glass2001} and
spatially structured collective behavior \cite{Tononi1998,Strogatz2005},
whose study requires elaborated methods. Recently, ordinal time series
analysis has received much attention because its tools present some
advantages like robustness against noise and computational efficiency.
Ordinal time series analysis is a particular form of symbolic analysis whose
`symbols'\ are ordinal patterns of a given length $L\geq 2$. This concept
was introduced by C. Bandt and B. Pompe in their seminal paper \cite%
{Bandt2002}, in which they also introduced permutation entropy as a
complexity measure of time series. Since then, ordinal time series analysis
has found a number of interesting applications in biomedical sciences,
physics, engineering, finance, statistics, etc.

Within the framework of ordinal symbolic dynamics, transcripts arise when
considering the relationship between coupled time series. Transcripts are
essentially ordinal patterns whose definition exploits the structure of the
permutation group. They were introduced in \cite{Monetti2009} and applied to
characterize the synchronization behavior of two coupled, chaotic
oscillators. This work was continued in \cite{Amigo2012}, where a general
concept of \textquotedblleft coupling complexity\textquotedblright\ was
given along with two complexity indices, $C_{1}$ and $C_{2}$, for its
quantification. Coupling complexity refers to the relationship among
dynamical system components; in general, it differs from the complexity of
the individual components or from their sum.

In this paper we approach some basic properties of the complexity index $%
C_{1}$ (which will be denoted hereafter just by $C$ or $C_{L}$) in a rather
didactic and intuitive way. For this reason our discussions and examples
will mainly address the cases of two or three coupled time series, shifting
the $N$-variate case to the Appendix. Examples also include the analysis of
real world data. In the last section, we extend the coupling complexity
index from random ordinal patterns to random ordinal pattern-valued
processes. The mathematically oriented reader will find in the Appendix a
number of technical facts, together with formal proofs of the properties
discussed in the main text.

\section{Transcripts}

We briefly describe the main concepts and some properties associated to
transcripts. Consider a time series $\{x_{t}\}_{t\geq 1}$ and a subsequence $%
x_{t}^{t+T(L-1)}=(x_{t},x_{t+T},\ldots ,x_{t+T(L-1)})$ of length $L$ and
time delay $T \geq 1$, extracted from $\{x_{t}\}_{t\geq 1}$. The \emph{%
ordinal pattern} $o(x_{t}^{t+T(L-1)})$ of length $L$ (or ordinal $L$%
-pattern) is defined as the rank-ordered indices of the components of $%
x_{t}^{t+T(L-1)}$. If, say,
\begin{equation*}
x_{t+Tj_{0}}<x_{t+Tj_{1}}<...<x_{t+Tj_{L-1}},
\end{equation*}%
then we write $o(x_{t}^{t+T(L-1)})=\left\langle
j_{0},j_{1}...,j_{L-1}\right\rangle $. Note that $\{j_{0},j_{1}...,j_{L-1}\}$
is a permutation of $\{0,1,...,L-1\}$. In case $x_{t+j_{r}}=x_{t+j_{s}}$, a
convention has to be used to order these two symbols. The whole sequence of
ordinal patterns extracted from $\{x_{t}\}_{t\geq 1}$ is known as the \emph{%
symbolic representation} of the time series.
\begin{figure}[tbp]
\begin{center}
\resizebox{0.85\columnwidth}{!}{
\includegraphics{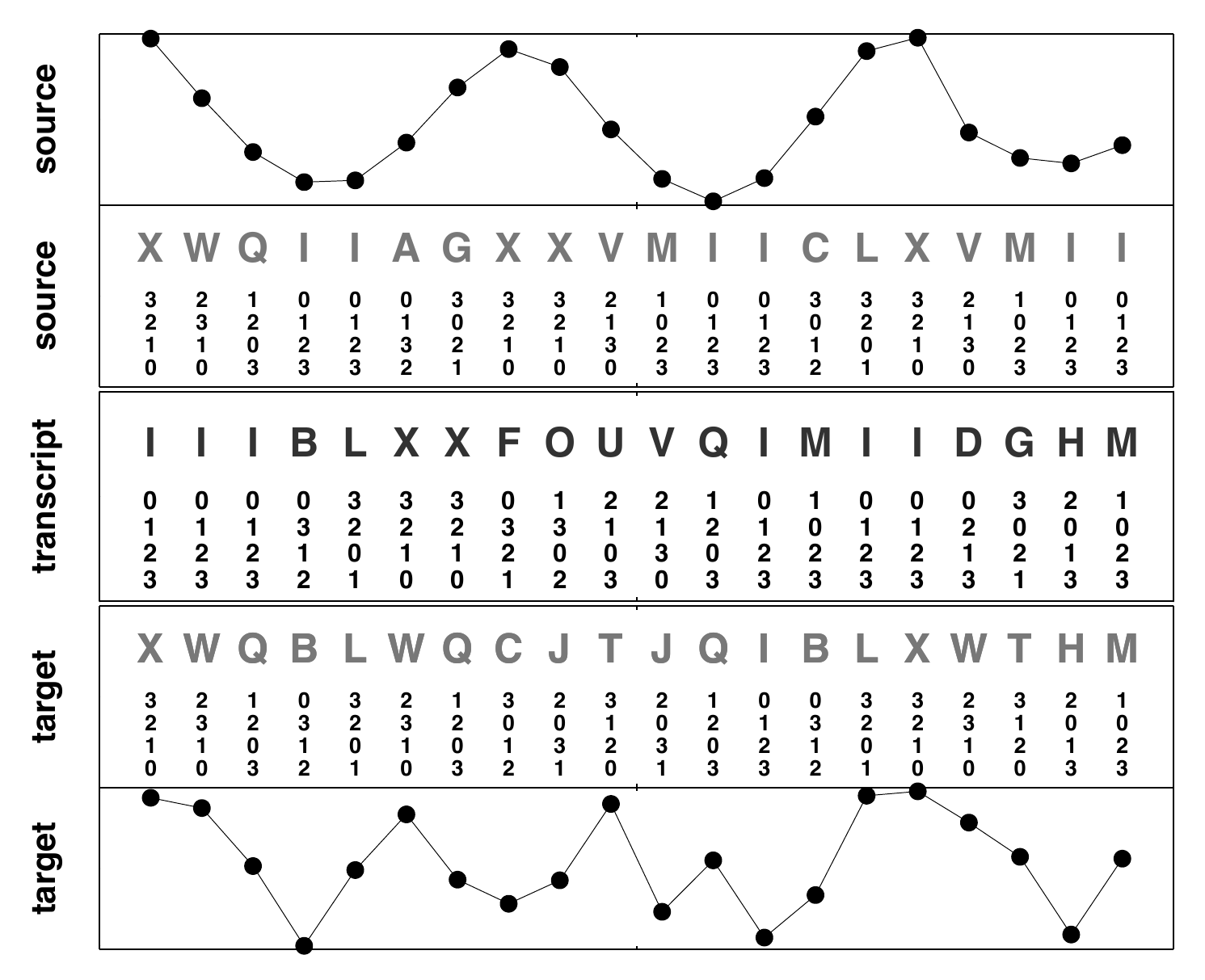}}
\end{center}
\caption{Transcription scheme for sequence length $L=4$. Symbols in light
grey (source and target) correspond to the symbolic representations and the
black symbols indicate the transcriptions that have to be applied to the
upper symbols (source) to obtain the lower ones (target).}
\label{fig:1}
\end{figure}

Ordinal patterns will be denoted by low case Greek letters throughout. Given
two symbols $\alpha $ and $\beta $ there always exists a unique symbol $\tau
_{\alpha ,\beta }=\tau $, in the following called \emph{transcript}, such
that the composition $\tau \alpha =\beta $. Specifically if $\alpha
=\left\langle j_{0},j_{1},\ldots ,j_{L-1}\right\rangle $ and $\tau
=\left\langle k_{0},k_{1},\ldots ,k_{L-1}\right\rangle $, then the action of
the symbol $\tau $ is defined as follows.%
\begin{equation}
\tau \alpha =\left\langle j_{k_{0}},j_{k_{1}},\ldots
,j_{k_{L-1}}\right\rangle .  \label{trdef}
\end{equation}
Figure \ref{fig:1} shows symbolic representations of the `source' and
`target' time series (light grey symbols) and the corresponding sequence of
transcripts (black symbols) for $L=4$. With the operation (\ref{trdef}), the
set of ordinal $L$-patterns forms a finite non-Abelian group of order $L!$
known as the \textit{symmetric group} $\mathcal{S}_{L}$ (i.e., the group of
permutations on $L$ elements). The identity permutation is%
\begin{equation*}
id=\left\langle 0,1,...,L-1\right\rangle ,
\end{equation*}%
and the inverse element is given by
\begin{equation*}
\alpha ^{-1}=o(j_{0},...,j_{L-1}).
\end{equation*}%
Then, $\tau _{\alpha ,\beta }=\beta \alpha ^{-1}$ and $\alpha =\tau _{\alpha
,\beta }^{-1}\beta $ are equivalent definitions of the transcript $\tau
_{\alpha ,\beta }$. In sum, the use of transcripts allows us to exploit the
structure of $\mathcal{S}_{L}$. For further properties of the transcripts,
see \cite{Monetti2009,Amigo2012}.

We now focus on the probability function of transcripts. Consider a source
and a target symbolic representations generated by the actual coupled
dynamics of the time series. Given a data source, the set of all feasible
ordinal $L$-patterns $\mathcal{S}^{1}=\{\alpha _{i}\}$ and $\mathcal{S}%
^{2}=\{\beta _{j}\}$ conform the state spaces of the source and the target
representations, respectively. Let $\Pr (\alpha )$ and $\Pr (\beta )$ be the
marginal probability functions, $\Pr (\alpha ,\beta )$ the joint probability
function, and
\begin{equation}
\Pr (\tau )=\sum_{\alpha ,\beta \in \mathcal{S}_{L}:\;\beta \alpha
^{-1}=\tau }\Pr (\alpha ,\beta )  \label{eq7}
\end{equation}%
the probability function of the transcripts. Thus, the entropy of the joint
probability function $\Pr (\alpha ,\beta )$ and the entropy of the
corresponding transcript probability function $\Pr (\tau )$ are given by
\begin{equation*}
H(\alpha ,\beta )=-\sum\limits_{\alpha ,\beta \in \mathcal{S}_{L}}\Pr
(\alpha ,\beta )\log \Pr (\alpha ,\beta ),
\end{equation*}

\noindent and

\begin{equation*}
H(\tau )=-\sum\limits_{\tau \in \mathcal{S}_{L}}\Pr (\tau )\log \Pr (\tau ),
\end{equation*}%
respectively.

The definition of transcript, Eq.~(\ref{trdef}), provides the algebraic
relationship between source and target ordinal patterns. It follows that,
given the triple $(\alpha ,\beta ,\tau )$, the knowledge of any pair of
symbols, i.e. $(\alpha ,\beta )$, $(\alpha ,\tau )$, or $(\beta ,\tau )$,
univocally determines the remaining symbol. We call this the\textit{\
uniqueness} \textit{property}. This important property implies \cite{bookIT}
that the corresponding joint entropies coincide, i.e.,
\begin{equation}
H(\alpha ,\beta )=H(\alpha ,\tau )=H(\beta ,\tau ).  \label{igual}
\end{equation}%
More generally, we mean by the uniqueness property the fact that different
sets of variables comprised of ordinal patterns and transcripts contain the
same information just because there is a 1-to-1 relation between any two of
them, allowing to recover the variables of one set from the variables of the
other. This property is used several times in the proofs of the Appendix.

\section{Coupling complexity}

Transcripts have triggered the development of information measures for the
assessment of the coupling relationship between system components. The
coupling complexity index is one of them. As its name indicates, this index
aims at quantifying the complexity of the interaction. Here, we consider
only one of two coupling complexity indices proposed in \cite{Amigo2012},
namely
\begin{equation}
C(\alpha ,\beta )=\min \{H(\alpha ),H(\beta )\}-\Delta (\alpha ,\beta ),
\label{theorem}
\end{equation}%
where
\begin{equation}
\Delta (\alpha ,\beta )=H(\alpha ,\beta )-H(\tau )  \label{delta}
\end{equation}%
is the \emph{information loss} of the transcription process \cite{Amigo2012}. According to \cite[Corollary 1]{Amigo2012}, $\Delta (\alpha ,\beta )\geq 0$. Whenever we want to underscore the length of the ordinal patterns used in the symbolic representation, we will write $C_{L}$, $\Delta _{L}$, etc.

We clearly observe in (\ref{theorem}) that the coupling complexity is
symmetric under the interchange of $\alpha $ and $\beta $. By means of Eq.~(%
\ref{igual}), $C$ can be written as
\begin{equation}
C(\alpha ,\beta )=\min \{I(\alpha ,\tau ),I(\beta ,\tau )\},  \label{C1}
\end{equation}%
where $I$ denotes the \textit{mutual information}. As mutual information is
a non-negative quantity, it follows $C(\alpha ,\beta )\geq 0$. Still a third
expression is
\begin{equation}
C(\alpha ,\beta )=H(\tau )-\max \{H(\alpha |\beta ),H(\beta |\alpha )\}.
\label{C2}
\end{equation}%
A generalization of (\ref{C1})-(\ref{C2}) to the multivariate case can be
found in the Appendix, Eqs. (\ref{1.2d})-(\ref{1.3}).

In order to deeper understand the meaning of $C$, we will discuss some cases
leading to extreme values. Assume two identical but otherwise arbitrary time
series. In this case, $\tau =id$ is the only existing transcript, thus $%
H(\tau )=0$. Since $H(\alpha ,\beta )=H(\alpha )=H(\beta )$, we conclude
that $C(\alpha ,\beta )=0$. As a second example, consider a symbolic
representation $\{\alpha _{t}\}_{t\geq 1}$ of a time series generated by
random $iid$ numbers along with another arbitrary one, $\{\beta
_{t}\}_{t\geq 1}$, independent of $\{\alpha _{t}\}_{t\geq 1}$. Since the
time series are independent, $H(\alpha ,\beta )=H(\alpha )+H(\beta )$. By
assumption, the probability function $\Pr (\alpha )$ is uniform and, owing
to the independence between $\alpha $ and $\beta $, the probability function
$\Pr (\tau )$ is uniform as well. Thus, $H(\tau )=H(\alpha )$ and $\min
\{H(\alpha ),H(\beta )\}=H(\beta )$, hence $C(\alpha ,\beta )=0$ once again.
We have shown with these two examples that the coupling complexity is a
property of the relationship between symbolic representations rather than a
characteristic of the individual components. It certainly vanishes if the
synchronization is rigid, but it also vanishes if the time series are
independent, provided that at least one of them is equidistributed.

\begin{figure}[tbp]
\begin{center}
\resizebox{0.5\columnwidth}{!}{
\includegraphics{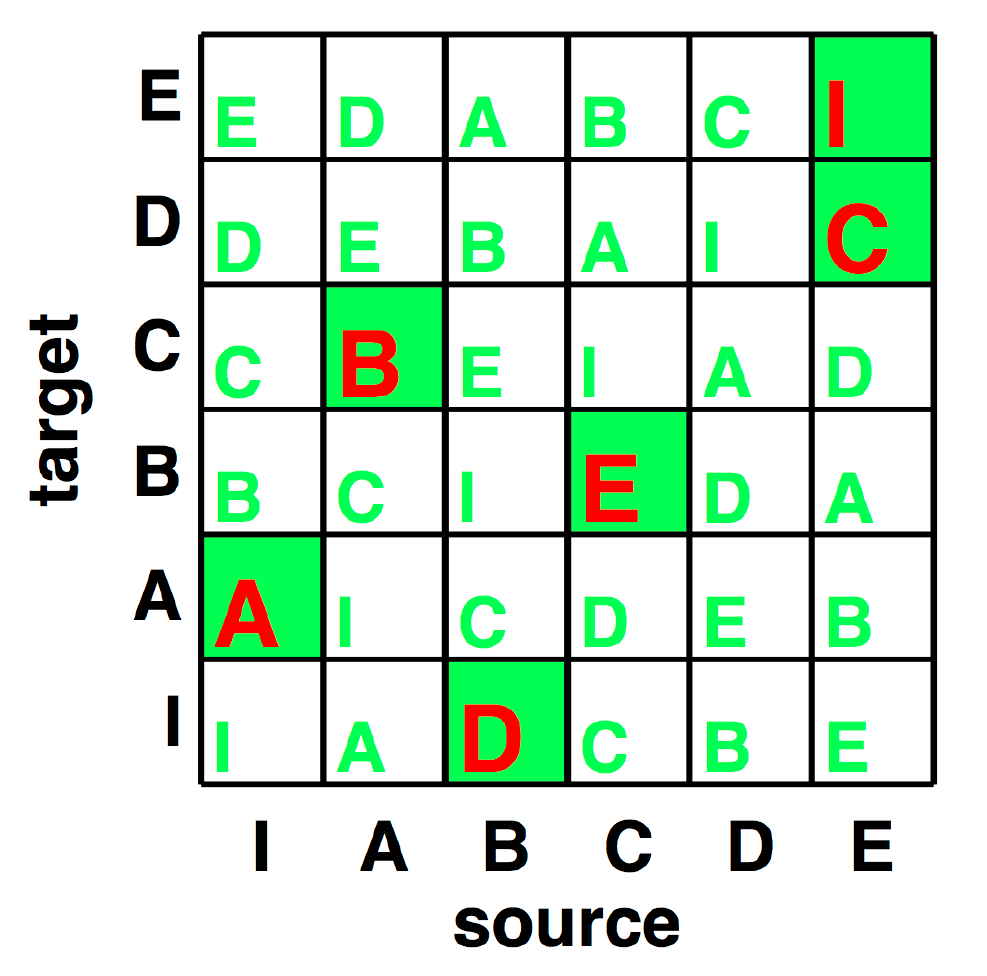}}
\end{center}
\caption{Transcription probability matrix for pattern length $L=3$ showing a
rare case with high coupling complexity. Symbols and transcripts were
denoted by the letters $I=\langle 0,1,2\rangle $, $A=\langle 0,2,1\rangle $,
$B=\langle 2,0,1\rangle $, $C=\langle 1,0,2\rangle $, $D=\langle
1,2,0\rangle $, and $E=\langle 2,1,0\rangle $. The matrix elements are the
transcripts which transform the `source symbol'\ into the `target symbol'.
In this example, only the transcripts in the shaded squares are supposed to
have positive and equal probability. Correspondingly, only some pairs of
symbols $(\protect\alpha ,\protect\beta )$ are realized.}
\label{fig:2}
\end{figure}
The example of Fig. \ref{fig:2} shows a transcription probability matrix for pattern length $L=3$,
where every transcript $\tau $ in a shaded square connects only one pair of
ordinal patterns $(\alpha ,\beta )$, where $\alpha ,\beta ,\tau \in
\{I,A,B,C,D,E\}$, while other connections are forbidden (more details in the
caption of Fig.~\ref{fig:2}). Furthermore, the allowed pairs occur all with
the same frequency. Thus, the knowledge of a transcript univocally
determines one pair of symbols, i.e. there is no loss of information in the
transcription process. Using this configuration, it is easy to calculate the
coupling complexity index. We first observe that $H(\tau )=H(\alpha ,\beta
)=\log 3!$, hence $\Delta (\alpha ,\beta )=0$. It is now clear why the
term $\Delta $ in Eq. (\ref{theorem}) was associated to the information loss
of the transcription, which is zero in this particular case. Consequently,
the coupling complexity reduces to $C(\alpha ,\beta )=\min \{H(\alpha
),H(\beta )\}=\log 3!-(2\log 2)/3!$, which is the maximum attainable complexity
value for $L=3$ and $\Delta _{L}=0$. A proof for $C_{L}(\alpha ,\beta )\leq $
$\log L!-(2\log 2)/L!$ if $\Delta _{L}=0$ can be found for $N$ symbolic
representations in the Appendix, Proposition \ref{MProp2}. This result
strongly suggests that $\log L!$ is not an optimal upper bound for the
coupling complexity index either in the bivariate nor in the multivariate
case.

\begin{figure}[tbp]
\begin{center}
\resizebox{0.75\columnwidth}{!}{
\includegraphics{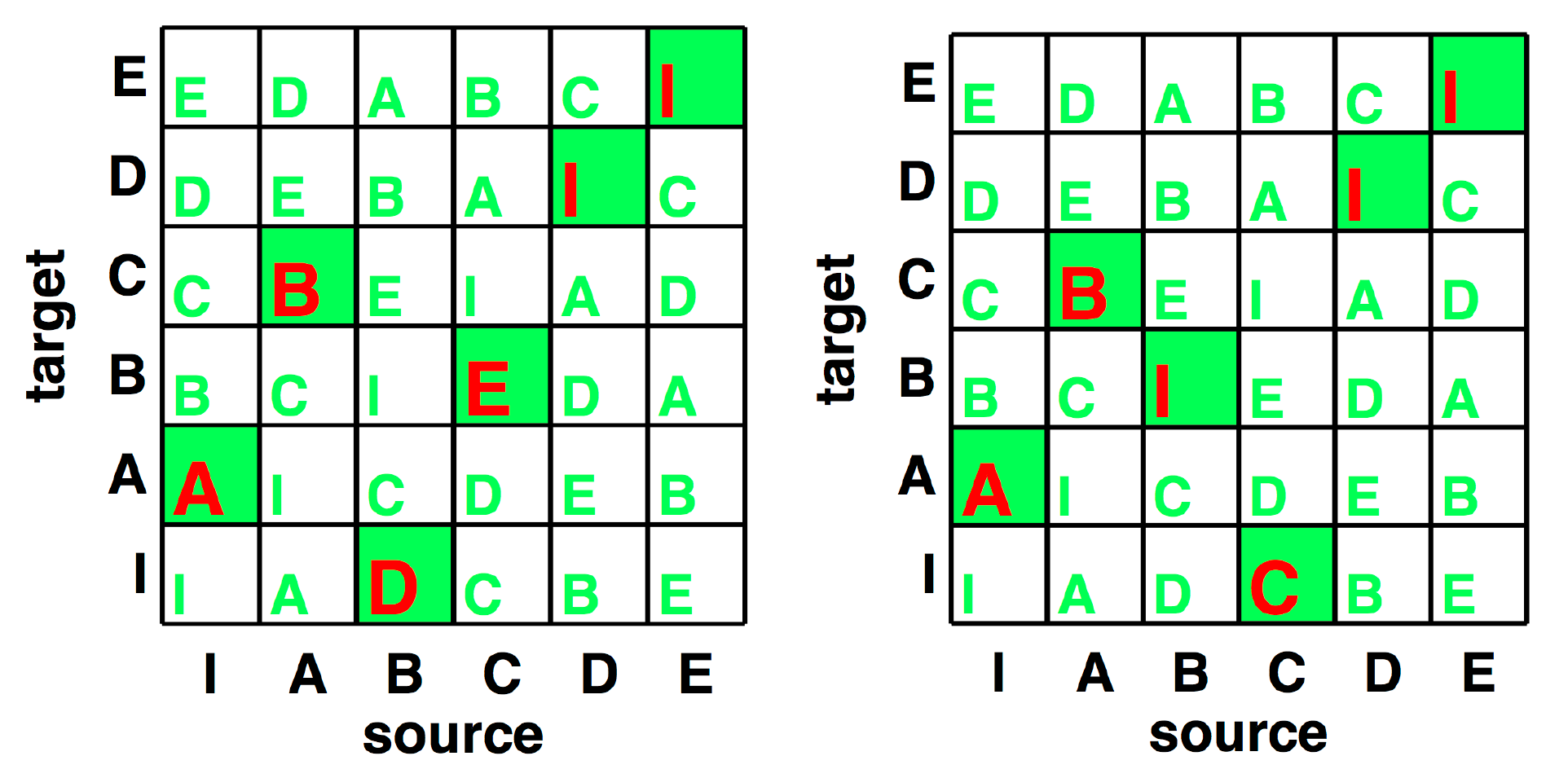}}
\end{center}
\caption{Transcription probability matrices for pattern length $L=3$
indicating two cases leading to different values of $C(\protect\alpha ,%
\protect\beta )$ but the same mutual information $I(\protect\alpha ,\protect%
\beta )$. Symbols are defined as in Fig.~\protect\ref{fig:2}. As before,
only the transcripts in the shaded squares are supposed to have positive and
equal probability.}
\label{fig3}
\end{figure}

Figure \ref{fig3} shows two examples again for the case $L=3$, which further clarify not only the
meaning of $\Delta (\alpha ,\beta )$ but also the difference between $%
C(\alpha ,\beta )$ and $I(\alpha ,\beta )$, the mutual information of the
symbolic representations. Since every row and every column of the
transcription probability matrices is occupied once and the probability is
supposed to be the same, we obtain $I(\alpha ,\beta )=\log 3!$ in both
cases. In order to calculate $C(\alpha ,\beta )$, we first note that both
marginal entropies are equal, and the joint entropy also equals the marginal
ones, i.e. $H(\alpha ,\beta )=H(\alpha )=H(\beta )$. Consequently, the
complexity index reduces to $C(\alpha ,\beta )=H(\tau )$ in both examples.
In the left panel of Fig. \ref{fig3}, we observe that only five transcripts
are realized, four of them having a 1-to-1 relationship with pairs of source
and target symbols, while the remaining one ($I$) is assigned to two
different pairs. This situation is reflected by the information loss of the
transcription, whose value is $\Delta (\alpha ,\beta )=(2\log 2)/3!$; the complexity
index amounts to $C(\alpha ,\beta )=\log 3!-(2\log 2)/3!$. Observe that this
value of $C$ is the same as we found for the transcription probability
matrix of Fig.~\ref{fig:2}, but both cases are distinguished by the
information loss $\Delta $. In the right panel of Fig. \ref{fig3}, the
situation is different, with only four transcripts realized, three of them
having a 1-to-1 relation with pairs of source and target symbols, the
remaining one ($I$) being assigned to three different pairs. Now, the
information loss of the transcription increases to $\Delta (\alpha ,\beta
)=(3\log 3)/3!$, while the complexity decreases to $C(\alpha ,\beta )=\log 3!-(3\log 3)/3!$.

The coupling complexity index can be generalized to the multivariate case by
means of the expression
\begin{eqnarray}
&&C(\alpha ^{1},\alpha ^{2},\ldots ,\alpha ^{N})  \label{c1hd} \\
&=&\min_{1\leq n\leq N}H(\alpha ^{n})-[H(\alpha ^{1},\alpha ^{2},\ldots
,\alpha ^{N})-H(\tau ^{1,2},\tau ^{2,3},\ldots ,\tau ^{(N-1),N})],  \notag
\end{eqnarray}%
where $\alpha ^{n}$ denotes the feasible symbols of the $n^{\mbox{th}}$ time
series and $\tau ^{(n-1),n}$ are the transcripts connecting the symbols $%
\alpha ^{n-1}$ and $\alpha ^{n}$, i.e., $\tau ^{(n-1),n}=\tau _{\alpha
^{n-1},\alpha ^{n}}$ for brevity. Similarly to the bivariate case, the
generalized coupling complexity is invariant under the interchange of the $%
\alpha ^{n}$'s. This property, which is a consequence of the uniqueness
property explained before, is proved in Proposition \ref{MProp3} of the
Appendix. For instance, consider three symbolic representations $\{\alpha
_{t}\}_{t\geq 1}$, $\{\beta _{t}\}_{t\geq 1}$, and $\{\gamma _{t}\}_{t\geq
1} $, and all possible transcripts $\{(\tau _{\alpha ,\beta })_{t}\}_{t\geq
1}$, $\{(\tau _{\alpha ,\gamma })_{t}\}_{t\geq 1}$, and $\{(\tau _{\beta
,\gamma })_{t}\}_{t\geq 1}$. The property of uniqueness warranties that $%
H(\tau _{\alpha ,\beta },\tau _{\alpha ,\gamma })=H(\tau _{\alpha ,\beta
},\tau _{\beta ,\gamma })=H(\tau _{\alpha ,\gamma },\tau _{\beta ,\gamma })$
and therefore the invariance of $C(\alpha ,\beta ,\gamma )$ (see Eq. (\ref%
{c1hd})) under permutations of its arguments.

Another consequence of the uniqueness property is the validity of Eq. (\ref%
{igual}) in higher dimensions, that is,
\begin{equation}
H(\alpha ^{1},\alpha ^{2},\ldots ,\alpha ^{N})=H(\alpha ^{n},\tau
^{1,2},\tau ^{2,3},\ldots ,\tau ^{(N-1),N})  \label{igualhd}
\end{equation}%
where $n\in \{1,\ldots ,N\}$. Using property (\ref{igualhd}) and the
following inequality (see for instance \cite{bookIT}),
\begin{equation}
H(Z|X,Y)\leq H(Z|X),  \notag
\end{equation}%
one can show under certain conditions the monotonicity of $C(\alpha
^{1},\ldots ,\alpha ^{N})$ with respect to the number of variables, that is
\begin{equation}
C(\alpha ^{1},...,\alpha ^{k},\ldots ,\alpha ^{N})\geq C(\alpha ^{1},...,%
\hat{\alpha}^{k},\ldots ,\alpha ^{N}),  \label{mono}
\end{equation}%
where $\hat{\alpha}^{k}$ means that $\alpha ^{k}$ has been omitted at that
position, and $N\geq 3$. For further details, see Proposition \ref{MProp4}
in the Appendix.

\begin{figure}[tbp]
\begin{center}
\resizebox{0.65\columnwidth}{!}{
\includegraphics{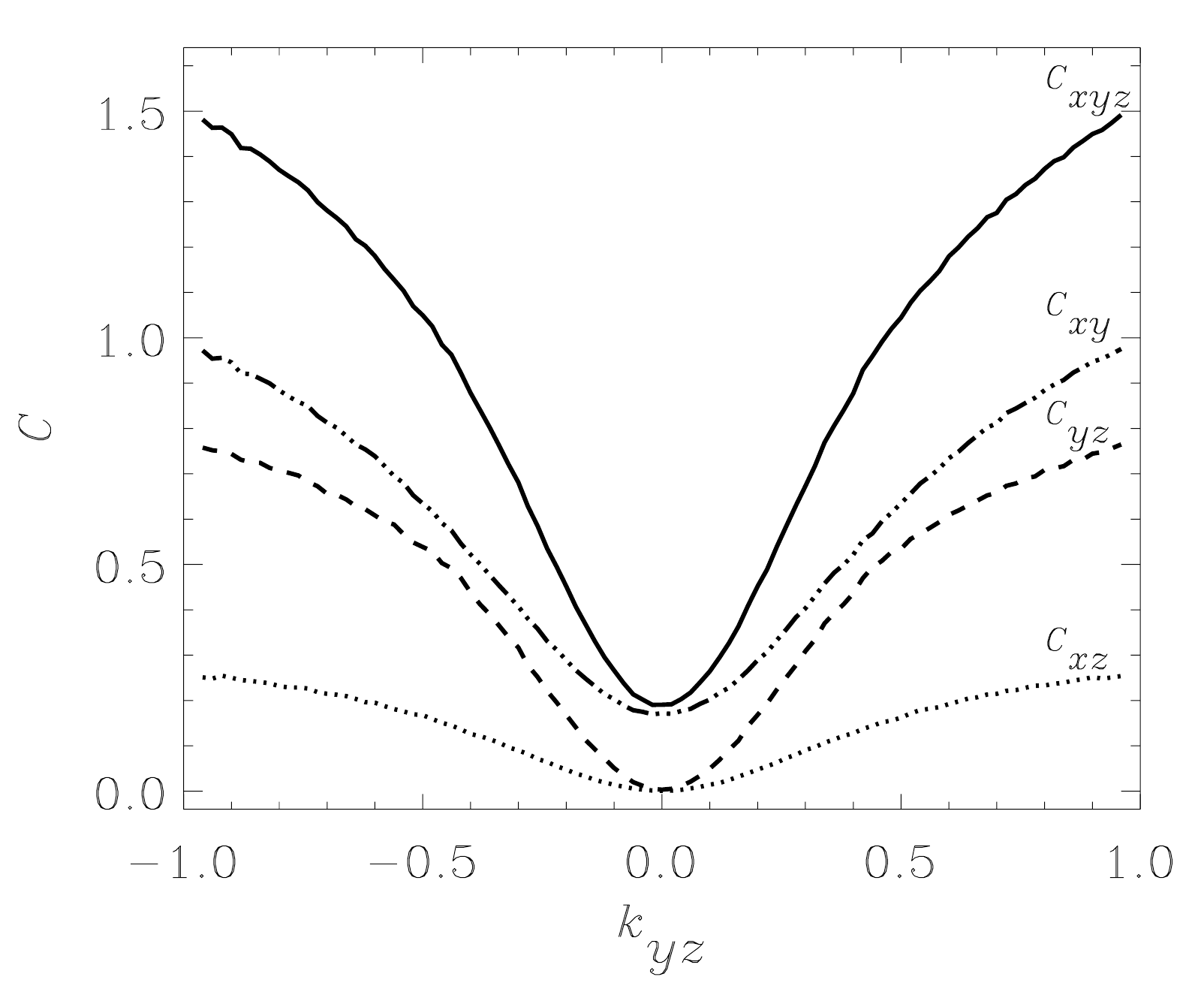}}
\end{center}
\caption{Coupling complexity indices evaluated for the 
coupled system defined in Eq. (\ref{xyz}).  The diagram shows the behaviour of the index of the three variables and the indices of all combinations of two variables
versus the coupling $k_{yz}$ for $k_{xy}=1$. The full, dot-dashed, dashed, and dotted
curves correspond to the complexities $C_{xyz}$, $C_{xy}$, $C_{yz}$, and $%
C_{xz}$, respectively. The symbolic representations were generated using
ordinal patterns of length $L=4$ and time delay $T=6$ for time series with $%
N=5\mbox{x}10^{5}$ data points.}
\label{fig4}
\end{figure}
As an example, consider the following three delay-coupled autoregressive
models,
\begin{eqnarray}
x(t+1) &=&ax(t)+k_{xy}y(t-\Lambda _{xy})+\eta ^{x}(t),  \label{xyz} \\
y(t+1) &=&by(t)+k_{yz}z(t-\Lambda _{yz})+\eta ^{y}(t),  \notag \\
z(t+1) &=&cz(t)+\eta ^{z}(t),\;\;\mbox{for}\;t\geq \max (\Lambda_{xy},\Lambda _{yz})  \notag
\end{eqnarray}
where $a=0.6$, $b=0.4$, $c=0.9$, $\Lambda _{xy}=3$, $\Lambda _{yz}=1$, and $%
\eta ^{x},\eta ^{y},\eta ^{z}$ are normal random numbers satisfying $%
\left\langle \eta ^{r}(t_{i})\eta ^{s}(t_{j})\right\rangle =\delta
_{rs}\delta _{ij}$ with $r,s\in \{x,y,z\}$. Here, we set the coupling
constant $k_{xy}=1$ and $k_{yz}\in (-1,1)$. The components of this system
are unidirectionally coupled in the form $z\rightarrow y\rightarrow x$, as
clearly indicated by Eq.~(\ref{xyz}). Denote by $\alpha ^{x},\alpha
^{y},\alpha ^{z}$ the symbols realized by the $x$-, $y$-, $z$-time series,
respectively. Figure~\ref{fig4} shows that $C_{xyz:}:=C(\alpha ^{x},\alpha
^{y},\alpha ^{z})\geq C(\alpha ^{r},\alpha ^{s})=:C_{rs}$, for every $r,s\in
\{x,y,z\}$, $r\neq s$, in agreement with Eq.~(\ref{mono}). It should be
noted that for $k_{yz}=0$, the components $y$ and $z$ become uncoupled but
the components $x$ and $y$ remain always coupled since $k_{xy}=1$.
Furthermore, due to the coupling structure of the model, the components $x$
and $z$ become uncoupled for $k_{yz}=0$ as well. Figure~\ref{fig4} shows
that for $k_{yz}=0$ both coupling complexities $C_{yz}$, and $C_{xz}$ almost
vanish. However, $C_{xy}$ remains significantly positive due to the
non-vanishing interaction between $x$ and $y$. It is worth noting that $%
C_{xyz}\simeq C_{xy}$ for $k_{yz}=0$, since the interaction $x$-$y$ is then
the only source of coupling complexity in the system. Let us point out that
the coincidence of $C_{xyz}$ with $C_{xy}$ in the uncoupled case ($k_{yz}=0$%
) is not valid in general (see Appendix, Proposition \ref{MProp5}, with $N=3$%
, $\alpha ^{1}=\alpha ^{x}$, $\alpha ^{2}=\alpha ^{y}$, $\alpha ^{3}=\alpha
^{z}$). However, in many cases the use of a suitable \textit{time delay} $T$
($T$ is expressed in samples of the time series, see Fig.~\ref{fig4}) may cause that the necessary conditions for the
validity of this property are fulfilled, thus clearly unveiling the coupling
structure of the dynamical system.

\section{Applications to real world data}

\begin{figure}[tbp]
\begin{center}
\resizebox{0.85\columnwidth}{!}{
\includegraphics{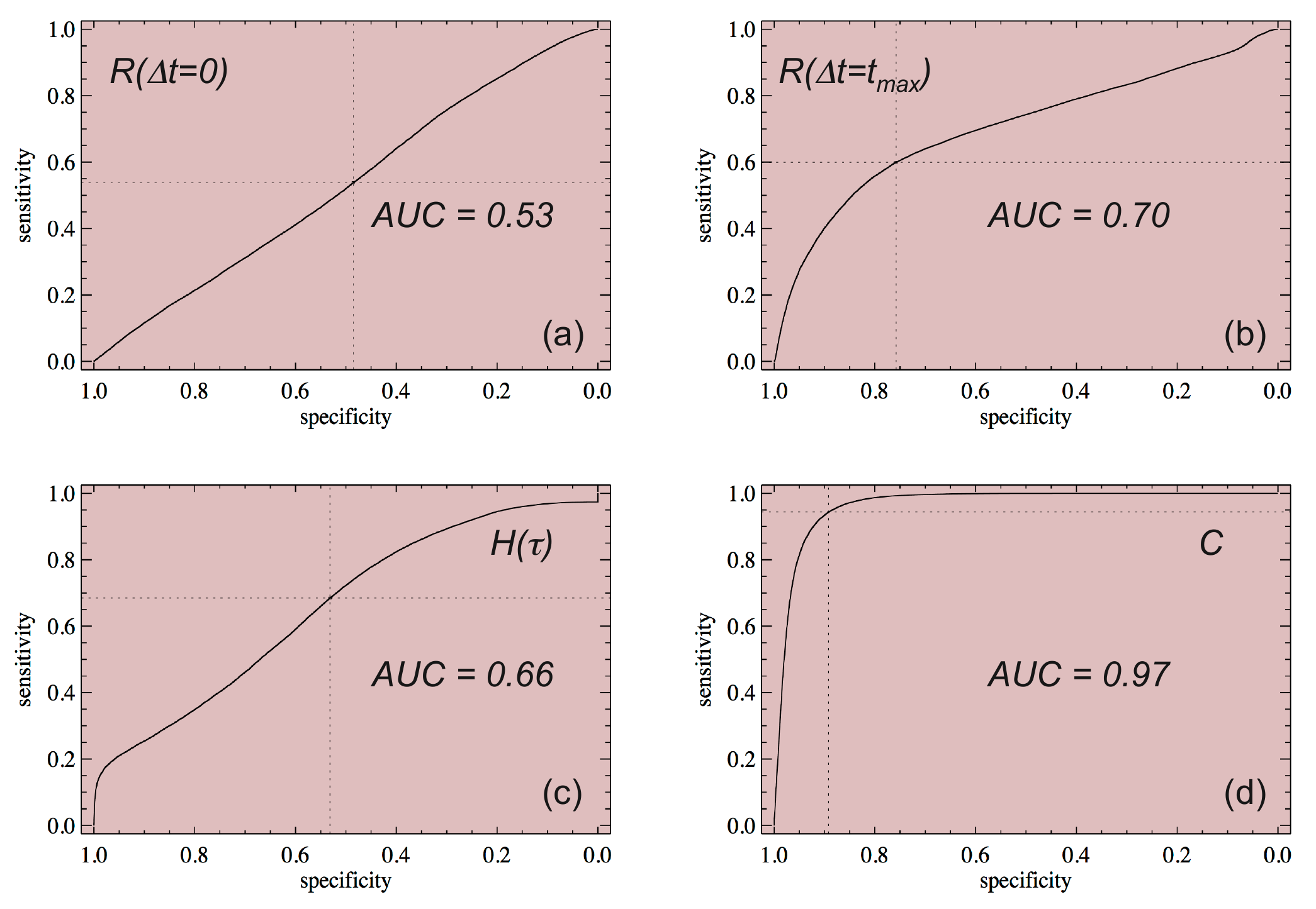}}
\end{center}
\caption{Receiver operator characteristic analysis to compare the
performance of different linear and non-linear measures to distinguish
pre-ictal from ictal states. The performance is quantified in terms of the
\textit{area under the curve} $AUC$, where $AUC=0.5$ is the expected performance
of a random classifier. All plots show sensitivity versus specificity. a)
Cross correlation $R$ evaluated at a time delay $\Delta t=0$. b) Maximum
value of the cross correlation. c) The entropy $H(\protect\tau )$ of the
transcripts. d) The coupling complexity $C$. More details in the text.}
\label{fig5}
\end{figure}

\begin{figure}[tbp]
\begin{center}
\resizebox{0.85\columnwidth}{!}{
\includegraphics{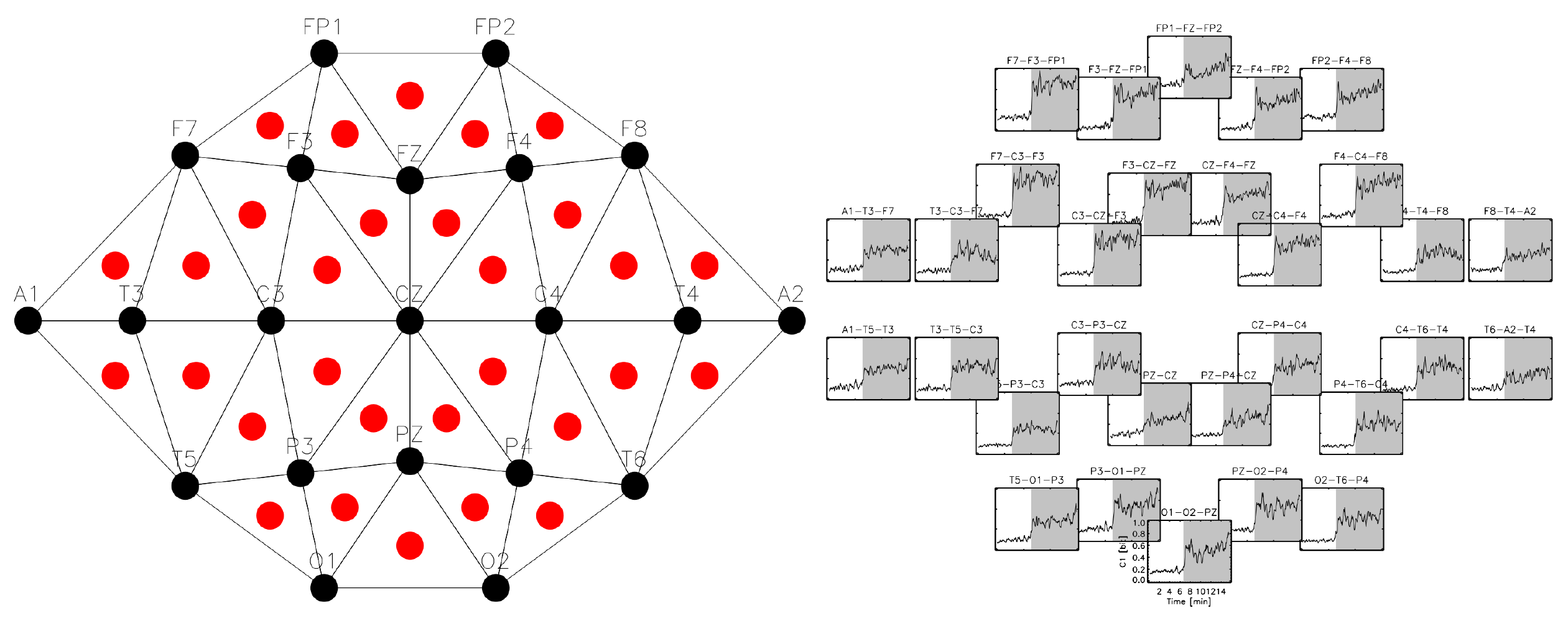}}
\end{center}
\caption{Left: Two dimensional diagram representing a two dimensional
projection of the position of the electrodes (black points, vertices of the
triangles) on the head, oriented as up-down $\rightarrow $
frontal-occipital, left-right $\rightarrow $ left ear - right ear. Right:
The time evolution of the complexity $C$ evaluated for every electrode
triple. The positions of the responses correspond to the grey points
centered on the triangles. All insets are displayed using the same scale,
thus they can easily be compared. The symbolic representations were generated
using ordinal patterns of length $L=3$ and time delay $T=0.048\sec$. More
details in text.}
\label{fig6}
\end{figure}

We analyze the electrical brain activity of an infant patient suffering from
\textit{frontal lobe epilepsy} (FLE). It should be remarked that it is not the
purpose of this work to perform a clinical study but to demonstrate the
applicability of the above presented methodology to an example of real world
data. A clinical study of the evolution of the brain electrical activity of
this infant during therapy has already been presented in Bunk et al. \cite%
{Bunk}. The authors compared the performance of a variety of synchronization
measures and found that the synchronization level is significantly increased
during the clinical manifestation of FLE, even in interictal periods. The
EEG recording was acquired during a time interval of 15 minutes at a
sampling rate of 250 Hz and a signal depth of 16 bits, and consists of 21
synchronously obtained time series. The positioning of the electrodes
followed that of the standardized 10-20-International System of Electrode
Placements (see the left panel of Fig.~\ref{fig6}). In \cite{Amigo2012}, we
demonstrated the applicability of the coupling complexity index $C$ to EEG
data in the context of epilepsy studies. By means of a sliding window
analysis considering all possible electrode pairs, we showed that $C$
increases in ictal periods, meaning that the relationship between different
EEGs becomes less random.

Here, we consider a simple task, namely, that of differentiating pre-ictal
and ictal states, in order to evaluate the performance of $C$ in comparison
with other linear and non-linear correlation measures. To this end, we
perform a sliding window analysis including all possible pairs of EEG
signals (210 pairs), using windows of size $w\simeq 16\sec $ and sampling
every $\Delta w\simeq 2\sec $. We have used patterns of length $L=4$ and a
time delay $T=0.048\sec $, thus every symbol has a time horizon of $\Delta
T=0.144\sec $. For every measure, all values in the pre-ictal and ictal
states are collected for all pairs, respectively. Then, the performance is
evaluated by means of a receiver operator characteristic analysis \cite%
{Swets1986}. Figure \ref{fig5} shows plots of the sensitivity versus the
specificity for the cross-correlations $R(\Delta t=0)$ and $R(\Delta
t=t_{\max })$, the entropy of the transcripts $H(\tau )$, and the coupling
complexity $C$. Here, $\Delta t$ denotes the time lag and $t_{\max }$ is the
time lag at the maximum of $R$. The \textit{sensitivity} is defined as $sen=%
\frac{Ntp}{Ntp+Nfn}$, where $Ntp$ is the number of true positives and $Nfn$
is the number of false negatives. Similarly, the \textit{specificity} is
given by the fraction $sp=\frac{Ntn}{Ntn+Nfp}$, where $Ntn$ is the number of
true negatives and $Nfp$ is the number of false positives. Thus, for a
random classifier, sensitivity and specificity amount about the same, and
the \textit{area under the curve} (AUC) becomes $AUC\simeq 0.5$. Figure \ref%
{fig5} (a) indicates that the cross-correlation $R(\Delta t=0)$ has a rather
poor performance, quite close to that of the random classifier. The
performance of the cross-correlation improves when considering the maximum $%
R(\Delta t=t_{\max })$, where now the brain state can be correctly predicted
in 70$\%$ of the cases. The entropy $H(\tau )$ has a performance rather
smaller than that of $R(\Delta t=t_{\max })$, but the best performance is
given by $C$ with an $AUC=0.97$. Thus, this example shows that $C$ captures
the main features essential to distinguish between these two brain states.

The application of $C$ to the multivariate case (Eq. (\ref{c1hd})) has to be
done with some care due to the curse of dimensionality. In fact, given $N$
time series the number of possible states for the joint process
(transcripts) is $L!^{N}$ ($L!^{N-1}$), respectively. Thus, one has to find
a suitable compromise between the number $N$ of time series to be analyzed
and the length $L$ of the ordinal patterns. We consider a trivariate
analysis of the above EEG data, where we have chosen neighboring triples of
electrodes as shown in the left panel of Fig~\ref{fig6}. Using $L=3$, a
rough estimate of the necessary data to perform such an analysis leads to $%
10\,\mbox{x}\,3!^{3}\simeq 2160$ data points, which is well below the window
length of $w=16\sec =4000$ data points used in this study. The right panel
of Fig.~\ref{fig6} shows for every electrode triple the time evolution of
the complexity $C$, where in every inset the white (grey) region corresponds
to the pre-ictal (ictal) state, respectively. Note that the positions of the
insets correspond to that of the central, grey points of the triangles on
the left panel of the same figure. We first observe in all insets that $C$
increases in the ictal state, in agreement with the results of the bivariate
analysis \cite{Amigo2012}. Furthermore, Fig.~\ref{fig6} shows that $C$
displays the strongest contrasts in the frontal region, thus supporting the
clinical diagnosis of \textit{frontal lobe epilepsy}.

\section{Coupling complexity rate}
Consider $N$ stationary random processes $\mathbf{X}^{n}=\{X_{t}^{n}\}_{t%
\geq1}$, $1\leq n\leq N$, where the identically distributed random variables
$X_{t}^{n}$ take values in a linearly ordered set. For each $t\geq1$ and $%
L\geq2$, set $\alpha_{t}^{n}=o(X_{t}^{n},...,X_{t+L-1}^{n})\in \mathcal{S}%
_{L}$. Therefore, each sequence $\mathbf{A}^{n}=\{\alpha_{t}^{n}\}_{t\geq1}$
is a stationary and identically distributed random process taking values in $%
\mathcal{S}_{L}$.

For each $n$, let $H(X^{n})$ be the entropy of the (identically distributed)
random variables $X_{t}^{n}$ and%
\begin{equation*}
h(\mathbf{X}^{n})=\lim_{t\rightarrow \infty }\frac{1}{t}%
H(X_{1}^{n},...,X_{t}^{n})
\end{equation*}%
the entropy rate of the process $\mathbf{X}^{n}$. The entropy $H_{L}(\alpha
^{n}):=H(o(X_{1}^{n},...,X_{L}^{n}))$, is called the permutation entropy of
order $L$ corresponding to the common probability distribution of the $%
\alpha _{t}^{n}$'s, and
\begin{equation}
h(\mathbf{A}^{n})=\lim_{L\rightarrow \infty }\frac{1}{L}H_{L}(\alpha
^{n})=:h^{\ast }(\mathbf{X}^{n})  \label{3.2}
\end{equation}%
is the permutation entropy rate of the process $\mathbf{X}^{n}$. It can be
proved that if the alphabet of $\mathbf{X}^{n}$ is finite, then $h^{\ast }(%
\mathbf{X}^{n})=h(\mathbf{X}^{n})$ \cite{Amigo2012a,Haruna2011}.

In (\ref{1.1a})-(\ref{1.1b}) we have defined the coupling complexity index $%
C_{L}(\alpha ^{1},...,\alpha ^{N})$ for random ordinal $L$-patterns $\alpha
^{1},...,\alpha ^{N}$. According to (\ref{3.2}), $H_{L}(\alpha ^{n})$ scales
linearly with $L$, and the same happens with $C_{L}(\alpha ^{1},...,\alpha
^{N})$ for $N$ fixed (see (\ref{c1hd})). Therefore, we define the
corresponding coupling complexity rate for symbolic processes as
\begin{equation}
C_{L}(\mathbf{A}^{1},...,\mathbf{A}^{N})=\underset{L\rightarrow \infty }{%
\lim \sup }\frac{1}{L}C_{L}(\alpha ^{1},...,\alpha ^{N}).  \label{3.3}
\end{equation}

This scenario can be extended to time series output by chaotic dynamical
systems (with sufficiently regular invariant measures). One important
difference between the randomly generated time series and the
deterministically generated ones is the existence in the latter case of
so-called \textit{forbidden patterns} (i.e., ordinal patterns that cannot
occur) under very general conditions \cite{Amigo2006,Amigo2008,Amigo2010}.
The bottom line is that the number of ordinal $L$-patterns actually observed
does not grow factorially but exponentially with $L$. To be specific, $%
\left\vert \{\alpha \in \mathcal{S}_{L}:\Pr (\alpha )>0\}\right\vert \propto
e^{Lh_{top}(f)}$, where $\propto $ stands for \textquotedblleft
asymptotically\textquotedblright , and $h_{top}(f)$ denotes the topological
entropy of $f$. Hence, if $h_{top}(f)<\infty $,
\begin{equation}
C_{L}(\alpha ^{1},...,\alpha ^{N})\leq \min_{1\leq n\leq N}H_{L}(\alpha
^{n})\leq Lh_{top}(f),  \label{3.4}
\end{equation}%
We conclude that if $\{\alpha _{t}^{n}\}_{t\geq 1}$ proceeds from an orbit
generated by a map $f:I^{d}\rightarrow I^{d}$, then%
\begin{equation*}
\frac{1}{L}C_{L}(\alpha ^{1},...,\alpha ^{N})\leq \frac{1}{L}\cdot
Lh_{top}(f)=h_{top}(f).
\end{equation*}%
Then $C(\mathbf{A}^{1},...,\mathbf{A}^{N})<\infty $ if $h_{top}(f)<\infty $.

\section{Conclusion}

In this paper we have briefly reviewed the concept of transcript and some of
its properties. We have also presented a detailed discussion of the coupling
complexity index, using examples that clarify its meaning. Furthermore, we
introduced a generalization of this index to $N$ time series and discussed
several useful properties via examples. A formal approach to the $N$-variate
coupling complexity index, containing mathematical proofs of some of its
properties, can be found in the Appendix below. As an application in higher
dimensions, we evaluated the coupling complexity index in a trivariate case
using biomedical data and found agreement with the results obtained in the
bivariate analysis. In the last section we introduced the coupling
complexity rate of symbolic random processes.

\begin{acknowledgement}
J.M.A. was financially supported by the Spanish
Ministry of Science and Innovation, Grant MTM2012-31698.
\end{acknowledgement}

\medskip

\appendix

\section{Properties of the $N$-variate coupling complexity index}


\label{appa} \setcounter{equation}{0} \numberwithin{equation}{section}

Two trivial properties of the transcripts are%
\begin{equation}
\tau _{\beta ,\alpha }=(\tau _{\alpha ,\beta })^{-1}  \label{0.1}
\end{equation}%
and%
\begin{equation}
\tau _{\beta ,\gamma }\tau _{\alpha ,\beta }=\gamma \beta ^{-1}\beta \alpha
^{-1}=\gamma \alpha ^{-1}=\tau _{\alpha ,\gamma }.  \label{0.2}
\end{equation}

Given the random variables $\alpha^{n}$, $1\leq n\leq N$, with outcomes in $%
\mathcal{S}_{L}$, then
\begin{align}
H(...,\alpha^{n},\alpha^{n+1},...) & =H(...,\alpha^{n},\tau_{\alpha
^{n},\alpha^{n+1}},...)=H(...,\alpha^{n},\tau_{\alpha^{n+1},\alpha^{n}},...)
\label{0.3a} \\
& =H(...,\tau_{\alpha^{n},\alpha^{n+1}},\alpha^{n+1},...)=H(...,\tau
_{\alpha^{n+1},\alpha^{n}},\alpha^{n+1},...)  \label{0.3b}
\end{align}
because any of the random variable pairs explicitly shown in (\ref{0.3a})-(%
\ref{0.3b}) can be determined from any other variable pairs.

For notational convenience we will also use the notations $\tau ^{n,n+1}$ or
$\tau (\alpha ^{n},\alpha ^{n+1})$ for $\tau _{\alpha ^{n},\alpha ^{n+1}}$,
and write, for example,
\begin{equation*}
H(\tau _{\alpha ^{1},\alpha ^{2}},...,\tau _{\alpha ^{N-1},\alpha
^{N}})\equiv H(\tau ^{1,2},...,\tau ^{N-1,N})\equiv H(\tau (\alpha
^{1},\alpha ^{2}),...,\tau (\alpha ^{N-1},\alpha ^{N})).
\end{equation*}%
If all ordinal $L$-patterns $\alpha ^{1},...,\alpha ^{N}$ are feasible, then
the $L!^{N-1}$ sets
\begin{eqnarray}
&&\Omega _{L}(\tau ^{1,2},...,\tau ^{N-1,N})  \label{0.5} \\
&=&\{(\alpha ^{1},...,\alpha ^{N})\in (\mathcal{S}_{L})^{N\text{ }}:\tau
^{1,2}\alpha ^{1}=\alpha ^{2},...,\tau ^{N-1,N}\alpha ^{N-1}=\alpha ^{N}\},
\notag
\end{eqnarray}%
build a partition of $(\mathcal{S}_{L})^{N\text{ }}$, with $\left\vert
\Omega _{L}(\tau ^{1,2},...,\tau ^{N-1,N})\right\vert =L!$, independently of
$N$. Actual time series may have symbolic representations with \textit{forbidden patterns} \cite{Amigo2006,Amigo2008,Amigo2010}\textbf{. }This may entail that
$\left\vert \Omega _{L}(\tau ^{1,2},...,\tau ^{N-1,N})\right\vert <L!$, or
even $\Omega _{L}(\tau ^{1,2},...,\tau ^{N-1,N})=\emptyset $ for some
choices of $\tau ^{1,2},...,\tau ^{N-1,N}$.

Consider throughout $N\geq 2$ random variables $\alpha ^{n}$, $1\leq n\leq N$%
, taking values in $\mathcal{S}_{L}$, sometimes called random ordinal $L$%
-patterns. Their\textit{\ coupling complexity index} was defined \cite%
{Amigo2012} as%
\begin{equation}
C_{L}(\alpha ^{1},...,\alpha ^{N})=\min_{1\leq n\leq N}H_{L}(\alpha
^{n})-\Delta _{L}\left( \alpha ^{1},...,\alpha ^{N}\right) ,  \label{1.1a}
\end{equation}%
where $H_{L}(\cdot )$ stands for the Shannon entropy of the random ordinal $%
L $-patterns appearing in the argument, and
\begin{equation}
\Delta _{L}\left( \alpha ^{1},...,\alpha ^{N}\right) =H_{L}(\alpha
^{1},...,\alpha ^{N})-H_{L}(\tau ^{1,2},...,\tau ^{N-1,N})  \label{1.1b}
\end{equation}%
is called the \textit{transcription information loss} \cite{Amigo2012}.
According to \cite[Corollary 1]{Amigo2012}, $\Delta (\alpha ,\beta )\geq 0$.
We dispense with the lower indices $L$ for the time being.

We are going to give an alternative expression for $C(\alpha ^{1},...,\alpha
^{N})$ that, in passing, will also prove that it is non-negative. First of
all, from%
\begin{equation}
H(\alpha ^{1},...,\alpha ^{N})=H(\alpha ^{k},\tau ^{1,2},...,\tau ^{N-1,N})
\label{1.1c}
\end{equation}%
it follows%
\begin{equation}
\Delta \left( \alpha ^{1},...,\alpha ^{N}\right) =H(\alpha ^{k}\left\vert
\tau ^{1,2},...,\tau ^{N-1,N}\right) ,  \label{1.1d}
\end{equation}%
$1\leq k\leq N$. Since $\Delta \left( \alpha ^{1},...,\alpha ^{N}\right) $
does not depend on $k$, the following result holds.

\bigskip

\begin{proposition}
\noindent \label{MProp1}$H\left( \alpha ^{i}\right\vert \tau ^{1,2},...,\tau
^{N-1,N})=H\left( \alpha ^{j}\right\vert \tau ^{1,2},...,\tau ^{N-1,N})$,
for every $1\leq i,j\leq N$.
\end{proposition}

\bigskip

Therefore,%
\begin{align}
C(\alpha^{1},...,\alpha^{N}) & =\min_{1\leq n\leq
N}H(\alpha^{n})-H(\alpha^{k}\left\vert \tau^{1,2},...,\tau^{N-1,N}\right)
\label{1.2a} \\
& =\min_{1\leq n\leq N}\left\{ H(\alpha^{n})-H(\alpha^{k}\left\vert
\tau^{1,2},...,\tau^{N-1,N}\right) \right\}  \notag \\
& =\min_{1\leq n\leq N}\left\{ H(\alpha^{n})-H(\alpha^{n}\left\vert
\tau^{1,2},...,\tau^{N-1,N}\right) \right\}  \notag \\
& =\min_{1\leq n\leq N}I(\alpha^{n};\tau^{1,2},...,\tau^{N-1,N})\geq0.
\label{1.2d}
\end{align}

Still other expression is%
\begin{align}
C(\alpha^{1},...,\alpha^{N}) & =H(\tau^{1,2},...,\tau^{N-1,N})-\left(
H(\alpha^{1},...,\alpha^{N})-\min_{1\leq n\leq N}H(\alpha^{n})\right)  \notag
\\
& =H(\tau^{1,2},...,\tau^{N-1,N})-\max_{1\leq n\leq N}H\left( \alpha
^{1},...,\hat{\alpha}^{n},...,\alpha^{N}\right\vert \left. \alpha^{n}\right)
,  \label{1.3}
\end{align}
where $\hat{\alpha}^{n}$ means that $\alpha^{n}$ has been omitted at that
position.

A trivial upper bound of $C(\alpha^{1},...,\alpha^{N})$ is $\log L!.$ The
next result suggests that this bound is not optimal.

\medskip

\begin{proposition}
\noindent \label{MProp2}If $\Delta \left( \alpha ^{1},...,\alpha ^{N}\right)
=0$, then
\begin{equation*}
C(\alpha ^{1},...,\alpha ^{N})=\min_{1\leq n\leq N}H(\alpha ^{n})\leq \log
L!-\frac{2}{L!}\log 2.
\end{equation*}
\end{proposition}

\medskip

\noindent\textit{Proof.} Suppose first $N\ =2$. In order that $\Delta
(\alpha^{1},\alpha^{2})=0$ there must be a 1-to-1 relation between each
realization $(\alpha_{i},\alpha_{j})$ of the pair $(\alpha^{1},\alpha^{2})$
and the ensuing transcript $\tau_{i,j}=\alpha_{j}(\alpha_{i})^{-1}$, i.e., $%
\left\vert \Omega_{L}(\tau)\right\vert =1$ for every $\tau\in\mathcal{S}_{L}$%
.

In turn, if given, say, any realization $\alpha_{i}$ of $\alpha^{1}$ there
would exist one and only one realization $\tau_{i,j}$ of $\tau^{1,2}$, then
this would fix also the realization $\alpha_{j}$ of $\alpha^{2}$ so that $%
H(\alpha^{1},\alpha^{2})=H(\alpha^{1})=H(\alpha^{2})$, and $C(\alpha
^{1},\alpha^{2})$ could reach its maximum $\log L!$. We prove by
contradiction that this cannot happen.

A 1-to-1 relation $\alpha_{i}\leftrightarrow\tau_{i,j}$ (or $\alpha
_{j}\leftrightarrow\tau_{i,j}=\tau_{j,i}^{-1}$ for that matter) means that
if $\alpha_{1},...,\alpha_{L!}$ is an enumeration of the ordinal patterns in
$\mathcal{S}_{L}$ (without restriction, $\alpha_{1}=\left\langle
1,2,...,L-1,L\right\rangle =:id$), then there is a permutation $\pi
:\{1,...,L!\}\rightarrow\{1,...,L!\}$ such that%
\begin{equation}
\alpha_{\pi(1)}(\alpha_{1})^{-1}=\tau_{1,\pi(1)},\;...,\;\alpha_{%
\pi(i_{0})}(\alpha_{i_{0}})^{-1}=id,\;...,\;\alpha_{\pi(L!)}(%
\alpha_{L!})^{-1}=\tau_{L!,\pi(L!)}  \label{1.5}
\end{equation}
is also an enumeration of $\mathcal{S}_{L}$. Note that $\pi(i_{0})=i_{0}$
but $\pi(i)\neq i$ for $i\neq i_{0}$ because, otherwise, the identity $id$
would be repeated on the list (\ref{1.5}).

We claim that such an enumeration is not possible. Indeed, from%
\begin{equation*}
\tau _{i,\pi (i)}\tau _{\pi (i),i}=\alpha _{\pi (i)}(\alpha _{i})^{-1}\alpha
_{i}(\alpha _{\pi (i)})^{-1}=id
\end{equation*}%
for every $i$, it follows that each of the $L!-1$ transcripts $\tau _{i,\pi
(i)}$ with $i\neq i_{0}$ can be paired with its inverse $(\tau _{i,\pi
(i)})^{-1}=\tau _{\pi (i),i}\neq \tau _{i,\pi (i)}$. Since their number is
odd, this is not possible. It follows that there is at least one $i\in
\{1,...,L!\}$ such that $\alpha _{i}$ is not realized by $\alpha ^{1}$,
hence $C(\alpha ^{1},\alpha ^{2})\leq H(\alpha ^{1})<\log L!$.

We conclude from the above argument that the maximum of $C\left( \alpha
^{1},\alpha^{2}\right) $ is reached when the would-be pairing $\alpha
_{i}\leftrightarrow\tau_{i,j}$ is minimally violated, i.e., there is only
one $\alpha_{i^{\prime}}$ with no correspondence $\tau_{i^{\prime},j}$ (thus
the pattern $\alpha_{i^{\prime}}$ is not realized by $\alpha^{1})$, and
there is only one $\alpha_{i^{\prime\prime}}$ with two correspondences $\tau
_{i^{\prime\prime},k}$ and $\tau_{i^{\prime\prime},h}$. Without restriction,
we assume that this happens for $i^{\prime}=L!$, and $i^{\prime\prime}=L!-1$%
, i.e., there is a map $\phi:\{1,...,L!-1\}\rightarrow\{1,...,L!\}$ with $%
\phi(i_{0})=i_{0}$, $\phi(i)\neq i$ for $i\neq i_{0}$, such that%
\begin{equation*}
\tau_{1,\phi(1)},\;...,\;\tau_{i_{0},\phi(i_{0})}=id,\;...,\;\tau
_{L!-1,\phi(L!-1)},\;\tau_{L!-1,K},
\end{equation*}
is also an enumeration of $\mathcal{S}_{L}$, where $K\in\{1,...,L!\}$ is
such that $K\neq\phi(i)$ for $i=1,...,L!-1$. To allow $H(\alpha^{1})$ to be
as great as possible, suppose furthermore that the transcripts are uniformly
distributed. Then
\begin{equation*}
H(\alpha^{1})=-(L!-2)\frac{1}{L!}\log\frac{1}{L!}-\frac{2}{L!}\log\frac{2}{L!%
}=\log L!-\frac{2}{L!}\log2,
\end{equation*}
$H(\alpha^{2})=\log L!$, and%
\begin{equation*}
C(\alpha^{1},\alpha^{2})=\min\{H(\alpha^{1}),H(\alpha^{2})\}=\log L!-\frac {2%
}{L!}\log2.
\end{equation*}

Finally, in the case $N>2$ extend consecutively the above rigid correlations
between $\alpha ^{1}$ with $\alpha ^{2}$ (which maximize $C(\alpha
^{1},\alpha ^{2})$) to $\alpha ^{3},...,\alpha ^{N}$. Then $\Delta (\alpha
^{1},...,\alpha ^{N})=0$, $H(\alpha ^{1})=...=H(\alpha ^{\left\lfloor
N+1\right\rfloor })=\log L!-\frac{2}{L!}\log 2$, $H(\alpha
^{2})=...=H(\alpha ^{\left\lfloor N/2\right\rfloor })=\log L!$, and $%
C(\alpha ^{1},...,\alpha ^{N})=\min_{1\leq n\leq N}H(\alpha ^{(n)})=\log L!-%
\frac{2}{L!}\log 2.$ $\square $

\medskip

Next we derive some basic properties of the complexity index.

\medskip

\begin{proposition}
\label{MProp3}\noindent \textbf{(}\textit{Invariance}\textbf{)}. $C(\alpha
^{1},...,\alpha ^{N})$ is invariant under permutations of $\alpha ^{1},$
..., $\alpha ^{N}$.
\end{proposition}

\medskip

\noindent\textit{Proof.} We claim that each random variable in the set $%
\mathcal{T}=\{\tau^{1,2},...,\tau^{N-1,N}\}$ is a function of random
variables in the set $\mathcal{T}_{\pi}=\{\tau^{\pi(1),\pi(2)},...,\tau
^{\pi(N-1),\pi(N)}\}$, and vice-versa, for any $\pi\in\mathcal{S}_{N}$. It
follows then
\begin{equation*}
H(\tau^{1,2},...,\tau^{N-1,N})=H(\tau^{\pi(1),\pi(2)},...,\tau^{\pi
(N-1),\pi(N)}),
\end{equation*}
and also $C(\alpha^{1},...,\alpha^{N})=C(\alpha^{\pi(1)},...,\alpha^{%
\pi(N)}) $ by the definition (\ref{1.1a})-(\ref{1.1b}).

In fact, we are going to prove the somewhat stronger result that $\mathcal{T}%
_{\pi }$ generates all $\tau ^{i,j}$, $1\leq i\neq j\leq N$, via
multiplication of transcripts sharing a common index. To this end, multiply
each transcript of $\mathcal{T}_{\pi }$ times the preceding one, that is, $%
\tau ^{\pi (n),\pi (n+1)}\tau ^{\pi (n-1),\pi (n)}=\tau ^{\pi (n-1),\pi
(n+1)}$ (see (\ref{0.2})), to obtain the set
\begin{equation*}
\mathcal{T}_{\pi }^{(2)}=\left\{ \tau ^{\pi (n),\pi (n+2)}:\;1\leq n\leq
N-2\right\} .
\end{equation*}%
$\mathcal{T}_{\pi }^{(2)}$ is the first member of a family of sets%
\begin{equation*}
\mathcal{T}_{\pi }^{(k)}=\left\{ \tau ^{\pi (n),\pi (n+k)}:\;1\leq n\leq
N-k\right\} ,
\end{equation*}%
$2\leq k\leq N-1$, recursively constructed as follows. Define $\mathcal{T}%
_{\pi }^{(1)}=\mathcal{T}_{\pi }$ and multiply all transcripts $\tau ^{\pi
(n),\pi (n+i)}\in \mathcal{T}_{\pi }^{(i)}$, $1\leq i\leq k-1$, times all
transcripts $\tau ^{\pi (n-k+i),\pi (n)}\in \mathcal{T}_{\pi }^{(k-i)}$,
sharing a common index $\pi (n)$ as indicated. According to (\ref{0.2}),
\begin{equation*}
\tau ^{\pi (n),\pi (n+i)}\tau ^{\pi (n-k+i),\pi (n)}=\tau ^{\pi (n-k+i),\pi
(n+i)}\in \mathcal{T}_{\pi }^{(k)}.
\end{equation*}
In particular, $\mathcal{T}_{\pi }^{(N-1)}=\left\{ \tau ^{\pi (1),\pi
(N)}\right\} $.

By construction, $\mathcal{T}_{\pi}^{(1)},\mathcal{T}_{\pi}^{(2)},...,%
\mathcal{T}_{\pi}^{(N-1)}$ build a partition of the set $\{\tau
^{\pi(n),\pi(m)}:1\leq n<m\leq N\}$ such that $\tau^{\pi(n),\pi(m)}\in%
\mathcal{T}_{\pi}^{(m-n)}$. The remaining transcripts are obtained by
inversion (see (\ref{0.1})).

Consider now an arbitrary transcript $\tau^{i,j}$, and let $\pi(a)=i$ and $%
\pi(b)=j$. Suppose that $a<b$, otherwise consider $\tau^{j,i}=(\tau
^{i,j})^{-1}$ instead$.$ Then $\tau^{i,j}=\tau^{\pi(a),\pi(b)}\in \mathcal{T}%
_{\pi}^{(b-a)}$. We conclude that any transcript $\tau^{i,j}$ (in
particular, those $\tau^{i,i+1}\in\mathcal{T}$) can be determined from $%
\mathcal{T}_{\pi}$ using algebraic operations (multiplications and
inversions) in the way explained above.

For the inverse relation, set $\beta ^{n}:=\alpha ^{\pi (n)}$ and proceed as
before, using this time the permutation $\pi ^{-1}$, to show that the
transcript $\tau (\beta ^{n},\beta ^{n+1})=\tau (\alpha ^{\pi (n)},\alpha
^{\pi (n+1)})$ can be recovered via algebraic operations from transcripts of
$\{\tau (\beta ^{(\pi ^{-1}(1))},\beta ^{(\pi ^{-1}(2))}),$...$,$ $\tau
(\beta ^{\pi ^{-1}(N-1)},\beta ^{\pi ^{-1}(N)})\}=\{\tau (\alpha ^{1},\alpha
^{2}),...,\tau (\alpha ^{N-1},\alpha ^{N})\}$. $\square $

\medskip

The next proposition shows that, under some provisos, $C$ is monotonous with
respect to the number of random variables.

\bigskip

\begin{proposition}
\noindent \label{MProp4}(\textit{Monotonicity}). Let $N\geq 3$.

\begin{description}
\item[(i)] If there is only one random variable $\alpha ^{n_{\min }}$ such
that $\min_{1\leq n\leq N}H(\alpha ^{n})=H(\alpha ^{n_{\min }})$, then $%
C(\alpha ^{1},...,\alpha ^{N})\geq C(\alpha ^{1},...,\hat{\alpha}%
^{k},...,\alpha ^{N})$ for all $k\neq n_{\min }$.

\item[(ii)] If there are at least two random variables with minimum entropy,
then $C(\alpha ^{1},...,\alpha ^{N})$ $\geq C(\alpha ^{1}$,...,$\hat{\alpha}%
^{k}$,...,$\alpha ^{N})$ for all $k=1,...,N$.
\end{description}
\end{proposition}

\medskip

\noindent \textit{Proof.} (i) If $k\neq n_{\min }$, then $\min_{n\neq
k}H(\alpha ^{n})=\min_{1\leq n\leq N}H(\alpha ^{n})=H(\alpha ^{n_{\min }})$.
Then, using (\ref{1.1d}) and Proposition \ref{MProp1},%
\begin{align*}
& C(\alpha ^{1},...,\alpha ^{N})-C(\alpha ^{1},...,\hat{\alpha}%
^{k},...,\alpha ^{N}) \\
& =H\left( \alpha ^{n_{\min }}\right\vert \tau ^{1,2},...,\tau
^{k-1,k+1},...\tau ^{N-1,N})-H\left( \alpha ^{n_{\min }}\right\vert \tau
^{1,2},...,\tau ^{N-1,N}) \\
& \geq 0
\end{align*}%
because the relative entropy $H(\alpha ^{n_{\min }}\left\vert ...\right) $
is greater the less conditioning variables.

(ii) In this case, one can always find for any $k\in \{1,...,N\}$ a random
ordinal $L$-pattern $\alpha ^{n}$ with $H(\alpha ^{n})=H(\alpha ^{n_{\min
}}) $ and $n\neq k$. Apply then (i). $\square $

\medskip

\begin{proposition}
\noindent \label{MProp5}If $\alpha ^{k}$ is independent of $\{\alpha
^{1},...,\hat{\alpha}^{k},...,\alpha ^{N}\}$ and $H(\alpha ^{k})=\log L!$,
then
\begin{equation*}
C(\alpha ^{1},...,\alpha ^{N})=C(\alpha ^{1},...,\hat{\alpha}^{k},...,\alpha
^{N}),
\end{equation*}%
where, in case $N=2$, we set $C(\alpha ^{1})=C(\alpha ^{2})=0$.
\end{proposition}

\medskip

\noindent \textit{Proof.} The case $N=2$ has to be dealt with separately.
If, say, $k=1$, then $\min \{H(\alpha ^{1}),$ $H(\alpha ^{2})\}=H(\alpha
^{2})$, $H(\alpha ^{1},\alpha ^{2})=H(\alpha ^{1})+H(\alpha ^{2})$ by
assumption, and%
\begin{eqnarray*}
\Pr (\tau ) &=&\sum_{\alpha ^{1},\alpha ^{2}:\tau \alpha ^{1}=\alpha
^{2}}\Pr (\alpha ^{1},\alpha ^{2})=\sum_{\alpha ^{1}}\Pr (\alpha ^{1})\Pr
(\tau \alpha ^{1}) \\
&=&\frac{1}{L!}\sum_{\alpha ^{1}}\Pr (\tau \alpha ^{1})=\frac{1}{L!},
\end{eqnarray*}%
hence $H(\tau )=\log L!=H(\alpha ^{1})$. It follows $C(\alpha ^{1},\alpha
^{2})=0=:C(\alpha ^{2}).$

Suppose $N\geq 3$. Since $C(\alpha ^{1},...,\alpha ^{N})\geq C(\alpha
^{1},...,\hat{\alpha}^{k},...,\alpha ^{N})$ by Proposition \ref{MProp4}
(ii), we need only to show that $C(\alpha ^{1},...,\alpha ^{N})\leq C(\alpha
^{1},...,\hat{\alpha}^{k},...,\alpha ^{N})$.

Using $\min_{1\leq n\leq N}H(\alpha ^{n})=\min_{n\neq k}H(\alpha ^{n})$ and
\begin{equation*}
H(\alpha ^{1},...,\alpha ^{N})=H(\alpha ^{k})+H(\alpha ^{1},...,\hat{\alpha}%
^{k},...,\alpha ^{N}),
\end{equation*}%
we obtain%
\begin{align*}
& C(\alpha ^{1},...,\alpha ^{N})-C(\alpha ^{1},...,\hat{\alpha}%
^{k},...,\alpha ^{N}) \\
& =-H(\alpha ^{1},...,\alpha ^{N})+H(\alpha ^{1,2},...,\tau ^{N-1,N}) \\
& +H(\alpha ^{1},...,\hat{\alpha}^{k},...,\alpha ^{N})-H(\tau
^{1,2},...,\tau ^{k-1,k+1},\tau ^{N-1,N}) \\
& =-H(\alpha ^{k})+H(\tau ^{1,2},...,\tau ^{N-1,N})-H(\tau ^{1,2},...,\tau
^{k-1,k+1},\tau ^{N-1,N}),
\end{align*}%
where%
\begin{equation*}
H(\tau ^{1,2},...,\tau ^{N-1,N})=H(\tau ^{n,k},\tau ^{1,2},...,\tau
^{k-1,k+1},\tau ^{N-1,N})
\end{equation*}%
for any $n\neq k$. Then%
\begin{align*}
& C(\alpha ^{1},...,\alpha ^{N})-C(\alpha ^{1},...,\hat{\alpha}%
^{k},...,\alpha ^{N}) \\
& =-H(\alpha ^{k})+H(\tau ^{n,k}\left\vert \tau ^{1,2},...,\tau
^{k-1,k+1},\tau ^{N-1,N}\right) \leq 0
\end{align*}%
%
because $H(\tau ^{n,k}\left\vert \tau ^{1,2},...,\tau ^{k-1,k+1},\tau
^{N-1,N}\right) \leq \log L!=H(\alpha ^{k})$. $\square $

\medskip

If all the random variables $\alpha ^{n}$ have a flat probability
distribution ($H(\alpha ^{1})=...=H(\alpha ^{1})=\log L!$), then, according
to Proposition \ref{MProp2}, $\Delta (\alpha ^{1},...,\alpha ^{N})>0$ and
\begin{equation}
C(\alpha ^{1},...,\alpha ^{N})=\log L!-\Delta (\alpha ^{1},...,\alpha
^{N})<\log L!\text{.}  \label{1.18}
\end{equation}%
The next proposition is a strengthening of (\ref{1.18}).

\medskip

\begin{proposition}
\noindent \label{MProp6}If all the random ordinal $L$-patterns $\alpha ^{n}$
are independent and uniformly distributed ($H(\alpha ^{1})=...=H(\alpha
^{1})=\log L!)$, then $C(\alpha ^{1},...,\alpha ^{N})=0$.
\end{proposition}

\medskip

\noindent\textit{Proof.} We have to prove (see (\ref{1.18})), that $%
\Delta(\alpha^{1},...,\alpha^{N})=\log L!$.

First of all,%
\begin{equation}
H(\alpha ^{1},...,\alpha ^{N})=H(\alpha ^{1})+...+H(\alpha ^{N})=N\log L!
\label{1.20}
\end{equation}%
because $\alpha ^{1},...,\alpha ^{N}$ are independent. Then (remember (\ref%
{0.5})),
\begin{align*}
\Pr (\tau ^{1,2},...,\tau ^{N-1,N})& =\sum_{(\alpha ^{1},...,\alpha ^{N})\in
\Omega _{L}(\tau ^{1,2},...,\tau ^{N-1,N})}\Pr (\alpha ^{1},...,\alpha ^{N})
\\
& =\sum_{(\alpha ^{1},...,\alpha ^{N})\in \Omega _{L}(\tau ^{1,2},...,\tau
^{N-1,N})}\Pr (\alpha ^{1})\cdots \Pr (\alpha ^{N}) \\
& =\sum_{(\alpha ^{1},...,\alpha ^{N})\in \Omega _{L}(\tau ^{1,2},...,\tau
^{N-1,N})}\frac{1}{(L!)^{N}}=\frac{1}{(L!)^{N-1}},
\end{align*}%
where we have used $\left\vert \Omega _{L}(\tau ^{1,2},...,\tau
^{N-1,N})\right\vert =L!$ in the last equality. We conclude that the
multivariate random variable $\tau ^{1,2},...,\tau ^{N-1,N}$ is also
uniformly distributed. Hence%
\begin{equation}
H(\tau ^{1,2},...,\tau ^{N-1,N})=\log (L!)^{N-1}=(N-1)\log L!.  \label{1.22}
\end{equation}

Finally, from (\ref{1.20}) and (\ref{1.22}),%
\begin{equation*}
\Delta (\alpha ^{1},...,\alpha ^{N})=H(\alpha ^{1},...,\alpha ^{N})-H(\tau
^{1,2},...,\tau ^{N-1,N})=\log L!\text{ }\square
\end{equation*}

\bigskip

\medskip
Received xxxx 20xx; revised xxxx 20xx. \medskip

\end{document}